\documentclass[
    ,final            
  ]
  {aipproc}
  
\layoutstyle{6x9}

\def\lsim{\mathrel{\rlap{\lower4pt\hbox{\hskip1pt$\sim$}}
    \raise1pt\hbox{$<$}}}         
\def\gsim{\mathrel{\rlap{\lower4pt\hbox{\hskip1pt$\sim$}}
    \raise1pt\hbox{$>$}}}         

\begin{document}

\title{Astrophysical Models of $r$-Process Nucleosynthesis: An Update}


\author{Yong-Zhong Qian}{
  address={School of Physics and Astronomy, University of Minnesota, Minneapolis, MN 55455, USA}
}

\begin{abstract}
 An update on astrophysical models for nucleosynthesis via rapid neutron capture, the $r$ process,
 is given. A neutrino-induced $r$ process in supernova helium shells may have operated up to metallicities 
 of $\sim 10^{-3}$ times the solar value. Another $r$-process source, possibly neutron star mergers, 
 is required for higher metallicities.
\end{abstract}

\maketitle


\section{Introduction}

Two major mechanisms are responsible for making heavy elements beyond Fe: the slow ($s$)
and rapid ($r$) neutron capture processes \cite{b2fh,agwc}. This contribution focuses on recent
developments in studies of the $r$ process. Extensive reviews on the subject can be found in e.g.,
\cite{thie,qian,arno}. See \cite{kapp} for a recent comprehensive review of $s$-process studies.

There are three major areas of $r$-process studies: (1) properties of extremely neutron-rich
nuclei far from stability that are involved in the process, (2) astrophysical sites and associated 
conditions that can be used to assess the feasibility of an $r$ process, and (3) frequencies of 
occurrences and other characteristics of potential $r$-process sources that are relevant for 
chemical evolution of galaxies. The second area is discussed in some detail below while the 
other two areas are mentioned only briefly.

The basic nuclear reactions involved in the $r$ process are neutron capture, photo-disintegration,
and $\beta$ decay. The crucial nuclear input for the $r$ process includes: (1) neutron capture rate 
coefficients $\langle v\sigma_{n,\gamma}(Z,A)\rangle$, which are averaged over thermal
distributions of neutron and nuclei, (2) photo-disintegration rates $\Lambda_{\gamma,n}(Z,A)$, 
which can be obtained from $\langle v\sigma_{n,\gamma}(Z,A-1)\rangle$ using detailed balance 
along with the neutron separation energy $S_n(Z,A)$ and relevant nuclear partition functions 
(e.g., \cite{thie}), and (3) $\beta$-decay rates $\lambda_\beta(Z,A)$. Here $(Z,A)$ 
stands for a nucleus with proton number $Z$ and mass number $A$. As the nuclei involved are
extremely neutron rich, branching ratios for emission of up to several neutrons following a $\beta$
decay also play an important role, especially during decay back to stability subsequent to freeze-out
of rapid neutron capture. If the $r$ process occurs in core-collapse supernovae, neutrino-induced
reactions are important as charged-current (CC) $\nu_e$ capture is equivalent to $\beta$ decay
and neutrino-induced neutron emission is efficient following both CC and neutral-current (NC) 
excitation \cite{qhlv}. If the $r$ process produces fissile nuclei, then rates for various fission
channels (e.g., spontaneous, neutron-induced, $\beta$-delayed, neutrino-induced)
and the corresponding mass distributions of fission fragments should also be included.
Only a small fraction of the nuclei involved in the $r$ process are accessible to experiments,
so the majority of the above nuclear input must be calculated from theory. As this input covers
a wide range of nuclear processes, a self-consistent theoretical framework for all the input is 
essential. Fortunately, the next generation of rare isotope beam facilities such as FRIB, RIKEN, 
and FAIR will provide crucial guidance to the theoretical efforts in addition to direct measurements 
of masses and $\beta$-decay rates of some key nuclei for the $r$ process.

By definition, an $r$ process occurs when the neutron capture rate of a typical nucleus involved
in the process exceeds its $\beta$-decay rate:
\begin{equation}
n_n\langle v\sigma_{n,\gamma}(Z,A)\rangle>\lambda_\beta(Z,A),
\end{equation}
where $n_n$ is the neutron (number) density at the astrophysical site. The quantity
$\langle v\sigma_{n,\gamma}(Z,A)\rangle$ depends on the temperature $T$ at the site
while $\lambda_\beta(Z,A)$ is essentially independent of $T$. The above 
equation can be used to give a crude estimate of $n_n$ required for an $r$ process:
\begin{equation}
n_n>10^{18}\left[\frac{10^{-17}\ {\rm cm}^3\ {\rm s}^{-1}}
{{\langle v\sigma_{n,\gamma}(Z,A)\rangle}}\right]
\left[\frac{\lambda_\beta(Z,A)}{10\ {\rm s}^{-1}}\right]\ {\rm cm}^{-3},
\label{eq-nn}
\end{equation}
where nominal values for $\langle v\sigma_{n,\gamma}(Z,A)\rangle$ and $\lambda_\beta(Z,A)$
have been used.

As a parcel of material undergoes $r$-processing, its $n_n$ and $T$ evolve with time $t$. 
The temperature $T$ is crucial in determining the path of the $r$ process. If $T$ is always
far below $10^9$~K, the $r$ process is cold and photo-disintegration is negligible. In this
case, the $r$-process flow is controlled by the competition between neutron capture and 
$\beta$ decay, and the path is populated dominantly by nuclei with
$\lambda_\beta(Z,A)\sim n_n\langle v\sigma_{n,\gamma}(Z,A)\rangle$.

If $T$ significantly exceeds $10^9$~K initially, a hot $r$ process occurs. 
For $T\sim 2\times 10^9$~K and $n_n\gsim 10^{20}$~cm$^{-3}$,
photo-disintegration is sufficiently fast to establish an equilibrium with neutron capture. In this 
$(n,\gamma)\rightleftharpoons(\gamma,n)$ equilibrium, the total
abundance of nuclei with a specific $Z$ is concentrated in a few so-called waiting-point nuclei, 
which are favored by the relevant $n_n$ and $T$ based mainly on their neutron 
separation energies $S_n(Z,A)$ and to a lesser extent, on their nuclear partition functions. 
As with any statistical equilibrium, the exact rates do not matter so long as they are fast enough 
for the equilibrium to hold. So the $r$-process path formed by the waiting-point nuclei
is specified by $n_n$, $T$, and $S_n(Z,A)$
but is insensitive to $\langle v\sigma_{n,\gamma}(Z,A)\rangle$ and $\Lambda_{\gamma,n}(Z,A)$
when $(n,\gamma)\rightleftharpoons(\gamma,n)$ equilibrium holds. The $\beta$ decay of
the waiting-point nuclei regulates the relative abundances for different $Z$ and
dominantly controls the progress of the $r$-process flow. However, when $n_n$ and $T$
drop below some critical values, $(n,\gamma)\rightleftharpoons(\gamma,n)$ equilibrium 
breaks down. This allows $\langle v\sigma_{n,\gamma}(Z,A)\rangle$ and 
$\Lambda_{\gamma,n}(Z,A)$ to play important roles in determining the final abundance 
pattern produced by a hot $r$ process.

For both cold and hot $r$ processes, the overall abundance pattern produced in an
individual astrophysical event, e.g., a core-collapse supernova, is a superposition of
abundance patterns produced in different parcels of material that have undergone 
different time evolution of $n_n$ and $T$. Astrophysical models of a hot $r$ process
are reviewed in the next section, followed by discussion of a cold
$r$ process induced by neutrino interaction in supernova helium shells.

\section{Astrophysical Models of a Hot $r$ Process}

In the absence of fission, the abundance
pattern produced in a parcel of material undergoing a hot $r$ process can be approximately
characterized by the corresponding average mass number $\langle A_r\rangle$. 
By mass conservation,
\begin{equation}
\langle A_r\rangle=\langle A_s\rangle+n/s,
\label{eq-nsr}
\end{equation}
where $\langle A_s\rangle$ is the average mass number of seed nuclei
that capture neutrons and $n/s$ is the number ratio of neutrons to seed nuclei
at the beginning of the $r$ process. So the conditions required for a hot $r$ process to 
produce nuclei with an average mass number $\langle A_r\rangle$ can be mapped to
those giving rise to the corresponding $\langle A_s\rangle$ and $n/s$. The viability of
a model for a hot $r$ process can then be assessed by how seed nuclei are produced
and how the neutron-to-seed ratio $n/s$ is determined before the actual $r$ process
occurs.

A generic sequence of evolution prior to a hot $r$ process can be illustrated by 
considering a parcel of material with an entropy $S\gsim 10$ in units of Boltzmann
constant per baryon that adiabatically expands from $T\gsim 10^{10}$~K.
The material at $T\gsim 10^{10}$~K is essentially composed of free nucleons
with $Y_n\approx 1-Y_e$, where $Y_n$ is the number fraction of neutrons and
$Y_e$ is the electron fraction (net number of electrons per nucleon). 
Given $S$ and $Y_e$, the nucleosynthesis during the expansion
can be calculated once the time evolution of $T$ is specified. A convenient parametric
form is $T(t)=T(0)\exp(-t/\tau_{\rm dyn})$, where $\tau_{\rm dyn}$ is a constant dynamic 
timescale. 

For $Y_e<0.5$, nearly all protons are assembled into $^4$He when $T$ decreases to 
$\sim 5\times 10^9$~K. At lower $T$, an $\alpha$ process occurs to burn neutrons and 
$^4$He \cite{wh92}. Here the role of $S$ is manifest.
Due to the absence of stable or long-lived nuclei at $A=5$ and 8,
the nuclear flow must go through the three-body reaction 
${{^4}{\rm He}}+{{^4}{\rm He}}+n\to{{^9}{\rm Be}}+\gamma$ to produce nuclei beyond
$^4$He. For $S>100$, there are a sufficient number of photons at or above
1.573~MeV, the threshold to break up $^9$Be, in the Planck distribution for
$5\times 10^9\gsim T\gsim 3\times10^9$~K. So the higher
$S$ is, the more inefficiently heavy nuclei are produced. In any case, 
charged-particle reactions effectively cease when $T$ drops to $\sim 3\times 10^9$~K
due to the Coulomb barrier. By then the inventory of heavy nuclei produced typically have 
$A\sim 60$--90. These nuclei act as seed nuclei to capture the remaining neutrons
during the subsequent $r$ process as $T$ continues to decrease.

\begin{figure}
  \includegraphics[angle=270,scale=0.45]{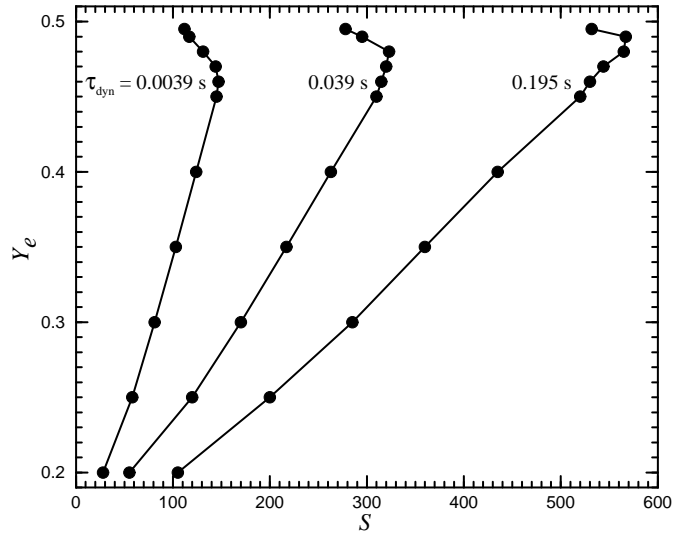}
  \caption{Combinations of $S$, $Y_e$, and $\tau_{\rm dyn}$ required for adiabatically expanding
  material to produce nuclei with $\langle A_r\rangle\approx 200$ through a hot $r$ process. See
  text and \protect\cite{hwq} for details.}
  \label{fig-sye}
\end{figure}

Clearly, there are an infinite number of combinations of $S$, $Y_e$, and $\tau_{\rm dyn}$
that can give the same $\langle A_r\rangle$ according to Eq.~\eqref{eq-nsr}. This can be
seen from the determination of $n/s$. Approximately
speaking, there are three different ways to produce a higher $n/s$: (1) a lower $Y_e$ to
increase the initial neutron abundance, (2) a higher $S$ to suppress production of seed
nuclei by keeping the abundance of $^9$Be low, and (3) a smaller $\tau_{\rm dyn}$ also 
to reduce seed production but by shortening the duration of the $\alpha$ process.
The $r$-process abundance pattern in the solar system is characterized by two peaks 
at $A\sim 130$ and 195, respectively. To produce the peak at $A\sim 195$ requires
$\langle A_r\rangle\approx 200$.
Figure~\ref{fig-sye} shows three contours in the $(S,Y_e)$-plane
that would all give $\langle A_r\rangle\approx 200$ for the corresponding values of 
$\tau_{\rm dyn}$ indicated. These results are taken from \cite{hwq} and can be used to
gauge the viability of an astrophysical environment to produce an $r$-process abundance 
peak at $A\sim 195$. Similar results can be found in \cite{mb97,frei}. Specific astrophysical 
models of a hot $r$ process are discussed below.

\subsection{Neutrino-driven winds from proto-neutron stars}

Core-collapse supernovae are intimately associated with neutrinos. Essentially all
the gravitational binding energy of a proto-neutron star formed in these events,
$\sim 3\times10^{53}$~erg, is radiated in $\nu_e$, $\bar\nu_e$,
$\nu_\mu$, $\bar\nu_\mu$, $\nu_\tau$, and $\bar\nu_\tau$ over $\sim 10$~s. A fraction
of the $\nu_e$ and $\bar\nu_e$ can be absorbed by the material immediately above
the proto-neutron star through the reactions
\begin{eqnarray}
\nu_e+n&\to&p+e^-,\label{eq-nuen}\\
\bar\nu_e+p&\to&n+e^+.\label{eq-nuebp}
\end{eqnarray}
During the $\sim 10$~s of neutrino emission, the proto-neutron star blows a wind as
the energy deposited by the above reactions continuously lifts shells of material out of 
its gravitational potential well. If reaction \eqref{eq-nuebp} proceeds more efficiently
than reaction \eqref{eq-nuen}, the wind material is driven neutron rich ($Y_e<0.5$)
\cite{qian93}. When this material reaches sufficiently large radii, these reactions
effectively cease due to small neutrino fluxes. Subsequently, the material approximately 
undergoes the generic evolution outlined above for adiabatic expansion with fixed
$S$, $Y_e$, and $\tau_{\rm dyn}$.

Earlier studies of nucleosynthesis in neutrino-driven winds based on the
supernova models of J.~R. Wilson showed that a hot $r$ process occurs
(e.g., \cite{meyer,woos94}). However, later studies \cite {witt,qian96,wana,thom,fisc,robe}
could not reproduce the high entropies in these models. Typically, $10\lsim S\lsim 100$,
$0.4\lsim Y_e\lsim 0.5$, and $0.01\lsim\tau_{\rm dyn}\lsim 0.1$~s were obtained.
For these conditions, the elements Sr, Y, and Zr with $A\sim 90$ are readily produced,
and for the most favorable among these conditions, 
the elements from Pd and Ag with $A\sim 110$ 
up to Te immediately below $A\sim 130$ can be produced (e.g., \cite{hwq}). However, all
these elements are produced in quasi-equilibrium of forward and reverse reactions 
involving photon, neutron, proton, and $\alpha$ particle but not by an $r$ process
(e.g., \cite{wh92}). In fact, heavy $r$-process elements with $A>130$, especially
those in the solar $r$-process abundance peak at $A\sim 195$, cannot be produced 
under these conditions (see Fig.~\ref{fig-sye}).

As $Y_e$ depends on the competition between reactions \eqref{eq-nuen} and
\eqref{eq-nuebp}, it is extremely sensitive to the difference between the emission
characteristics of $\nu_e$ and $\bar\nu_e$ \cite{qian93}. 
In contrast, $S$ and $\tau_{\rm dyn}$
are not sensitive to this difference as they depend on the net heating by both
$\nu_e$ and $\bar\nu_e$. Recent calculations including more neutrino transport 
processes showed that neutrino-driven winds are always proton rich ($Y_e>0.5$) 
\cite{fisc,hude}. This gives rise to a new process of nucleosynthesis, the $\nu p$
process \cite{prue,froh}, which produces elements beyond the Fe group
mostly via proton capture. Without neutrinos, those ``waiting-point'' nuclei in
$(p,\gamma)\rightleftharpoons(\gamma,p)$ equilibrium must $\beta$ decay
before the nuclear flow can proceed further (cf. a hot $r$ process) and a
waiting-point nucleus with a very slow $\beta$-decay rate effectively terminates
the flow. The flow proceeds much further in the presence of an 
intense $\bar\nu_e$ flux because the neutron produced by reaction 
\eqref{eq-nuebp} can enable an $(n,p)$ reaction that
has the same effect as but is much faster than $\beta$ decay. Therefore,
the $\nu p$ process can produce elements far beyond the Fe group.
It was shown that the elements from Sr, Y, and Zr to Pd and Ag with
$A\sim 90$--110 can be produced in both proton-rich and neutron-rich winds
\cite{arco}. So while neutrino-driven winds are unlikely to produce heavy
$r$-process elements with $A>130$, they are an important source of the
above elements. Further, they served as a detailed case study of
novel processes of nucleosynthesis that were not envisioned in the classical 
papers of Burbidge et al. \cite{b2fh} and Cameron \cite{agwc}
on the origin of the elements.

\subsection{Shocked surface layers of O-Ne-Mg cores}

Stars with masses $M\gsim 11M_\odot$ develop Fe cores and those 
with $8\lsim M<11M_\odot$ develop O-Ne-Mg cores before they undergo
core collapse (e.g., \cite{whw}). These two groups of stars differ greatly
in structure as the density above an O-Ne-Mg core falls off much more 
drastically than that above an Fe core \cite{nomo}. 
So far O-Ne-Mg core-collapse
supernovae are the only case for which neutrino heating
[mostly via reactions \eqref{eq-nuen} and \eqref{eq-nuebp} as in the case of 
neutrino-driven winds] has been demonstrated to result in a successful
explosion (e.g. \cite{kita}). 

The surface layers of an O-Ne-Mg core are slightly neutron rich due to
processing by pre-supernova nuclear burning. The evolution of such a layer
subsequent to being struck by the supernova shock 
again approximately follows adiabatic expansion
with fixed $S$, $Y_e$, and $\tau_{\rm dyn}$. Due to
the steep density gradient above the O-Ne-Mg core,
the shocked material may expand rapidly with $\tau_{\rm dyn}\sim 10^{-3}$~s. 
This along with $S\sim 100$ and $Y_e<0.5$ (see Fig.~\ref{fig-sye})
would enable a hot $r$ process to produce
nuclei with $\langle A_r\rangle \approx 200$ \cite{ning}. However, current
models of O-Ne-Mg core collapse based on Nomoto's models of the
associated supernova progenitors \cite{nomo}
only give a weak explosion, for which 
the shocked surface layers of the core expand with $\tau_{\rm dyn}\gg 10^{-3}$~s 
and $S\ll 100$ \cite{jank}. As these layers are not extremely neutron rich,
no hot $r$ process can occur under the conditions obtained in current models.
It remains to be seen if more favorable conditions can be obtained for new models
of stars that develop O-Ne-Mg cores.

\subsection{Winds from accretion disks of black holes}

Figure~\ref{fig-sye} shows that for $\tau_{\rm dyn}/S^3\sim 2\times 10^{-7}$~s
with $25\lsim S\lsim 100$ and 
$3.9\times 10^{-3}\lsim\tau_{\rm dyn}\lsim0.195$~s, nuclei with 
$\langle A_r\rangle \approx 200$ can be produced by a hot $r$ process
for $Y_e\sim 0.2$. The above ranges of $S$ and $\tau_{\rm dyn}$ 
occur in neutrino-driven winds from proto-neutron stars.
However, $Y_e\sim 0.2$ cannot be obtained in these winds without 
invoking new physics such as sterile neutrinos \cite{mcla}. 
Therefore, it is important to explore alternative candidate environments for 
a hot $r$ process where such low $Y_e$ values may be obtained more naturally
along with similar $S$ and $\tau_{\rm dyn}$ to those in neutrino-driven winds.

Some Fe-core collapse supernovae are expected to produce black holes
rather than neutron stars. An accretion disk typically forms around the
black hole. Such disks can also form in mergers of two neutron stars or
of a neutron star with a black hole.
The disk material close to the black hole is hot and dense and emits
neutrinos similarly to the surface layers of a proto-neutron star. In addition, 
this material may be very neutron rich as a result of the reaction $e^-+p\to n+\nu_e$. 
A fraction of this material is ejected in winds due to heating by dynamic processes 
in the disk or by neutrino reactions \eqref{eq-nuen} and \eqref{eq-nuebp}.
Parametric models suggest that adequate conditions, in
particular low $Y_e$ values, may be obtained in these winds 
for a hot $r$ process to occur \cite{pwh,smh,surm,waja}. It remains to be seen
if such conditions can be derived in self-consistent models of the accretion disk 
and the associated winds. It is especially important that detailed neutrino transport 
in the disk be carried out similarly to that in modern core-collapse supernova models.

\subsection{Neutron star mergers}

Material of low $Y_e$ stocked up inside a neutron star can be released by tidal force
during mergers of two neutron stars or of a neutron star with a black hole.  Once
released from the strong gravitational hold of its parent neutron star, such material
undergoes decompression, which is accompanied by heating due to $\beta$ decay.
The possibility of a hot $r$ process during the expansion of decompressed 
neutron-star matter was discussed in the pioneering work of Lattimer et al. \cite{latt}
and more recently in \cite{ross,gori}. An interesting feature of this $r$-process model
is that the nuclear flow typically reaches fissile nuclei and becomes cyclic as the 
fission fragments capture neutrons to replenish the flow. This fission cycling is
expected to produce a robust abundance pattern at $A>130$, which appears to be
in accord with observations of metal-poor stars in the Galactic halo 
but with some notable exceptions (e.g., \cite{sned,hond}).

A potential problem with $r$-process models based on neutron star mergers is that 
these events occur $\sim 10^3$ times less frequently than core-collapse supernovae 
in the Galaxy. Were neutron star mergers a major source for heavy $r$-process elements 
with $A>130$, stars with the earliest such enrichments would have been enriched with 
high Fe abundances by the numerous core-collapse supernovae that had already occurred
prior to the contributing $r$-process event \cite{qian00,arga}. 
This is in contradiction to observations of a number of metal-poor stars
that are enriched in heavy $r$-process elements but have 
[Fe/H]~$\equiv\log({\rm Fe/H})-\log({\rm Fe/H})_\odot\lsim -3$ (e.g., \cite{sned}).
It remains to be seen if more detailed models of Galactic chemical evolution 
may resolve the above difficulty.

\section{Neutrino-Induced Nucleosynthesis in Supernova Helium Shells: A Cold $r$ Process}

The notion of a unique type of sources that produce heavy $r$-process elements
with $A>130$ over the entire history of the Galaxy has some appeal but is far from 
required by observations (e.g., \cite{hond,wbg}). As discussed above, winds from black hole 
accretion disks and decompressed ejecta from neutron star mergers could both be production 
sites for such elements. The difficulty of neutron star mergers to account for observed
enrichments of heavy $r$-process elements at [Fe/H]~$\lsim -3$ leads to the question
whether other sources could have operated at such low metallicities. In fact, such a source
was proposed by Epstein, Colgate, and Haxton (ECH) \cite{ech}. In their model, NC neutrino
reactions on $^4$He nuclei produce neutrons for an $r$ process
via ${^4{\rm He}}(\nu,\nu n){^3{\rm He}}(n,p){^3{\rm H}}$
or ${^4{\rm He}}(\nu,\nu p){^3{\rm H}}$ followed by ${^3{\rm H}}({^3{\rm H}},2n){^4{\rm He}}$.
This occurs in helium shells of metal-poor core-collapse supernovae, where neutrons are
left alone by the predominant $^4$He nuclei but captured by the Fe nuclei acquired at the birth 
of the supernova progenitors. The helium shells have 
$T\sim 10^8$~K, and therefore, the ECH mechanism is a cold $r$ process.
The neutron density for this mechanism depends on the competition between production by 
neutrino reactions and capture by Fe and other nuclei. Consequently, $n_n\gsim 10^{18}$~cm$^{-3}$
required for an $r$ process [see Eq.~\eqref{eq-nn}] can only be obtained at low metallicities.
The ECH mechanism was critiqued in \cite{nup} as viable only for low-mass metal-poor 
supernovae, where helium shells are located at relatively small radii to ensure sufficiently
large neutrino fluxes. Follow-up studies were carried out in e.g., \cite{nady}. 

Recently, it was shown that a modified ECH mechanism could operate in core-collapse supernovae
with [Fe/H]~$\lsim -3$ over a range of progenitor masses \cite{bhq}. For a $15\,M_\odot$
progenitor with [Fe/H]~$\sim -4.5$ \cite{whw}, a cold $r$ process occurs in the outer helium
shells at radii of $\sim 10^{10}$~cm, where the abundances of $^{12}$C and $^{16}$O nuclei
produced by pre-supernova helium burning are too low to cause significant neutron capture.
The CC reaction ${^4{\rm He}}(\bar\nu_e,e^+n){^3{\rm H}}$ turns out to be
the dominant source of neutrons because the temperature is too low for the reaction
${^3{\rm H}}({^3{\rm H}},2n){^4{\rm He}}$ to be effective \cite{bhq}. Further, the CC reaction
rate is greatly enhanced when the energy spectra of $\bar\nu_\mu$ and $\bar\nu_\tau$ are
much harder than that of $\bar\nu_e$ at emission and an inverted neutrino mass hierarchy
enables $\bar\nu_e\rightleftharpoons\bar\nu_{\mu,\tau}$ transformation below the outer 
helium shells \cite{bhq}. More systematic studies of the modified ECH mechanism are needed
to assess its contribution to Galactic chemical evolution. It is conceivable that this mechanism
operated in the early Galaxy until [Fe/H]~$\sim -3$ and neutron star mergers provided
the subsequent $r$-process enrichments.


\begin{theacknowledgments}
  I thank Projjwal Banerjee, Wick Haxton, and Alexander Heger for collaboration on
  neutrino-induced nucleosynthesis in core-collapse supernovae, and
  Jie Meng and Baohua Sun for collaboration on nuclear physics input for the $r$ process. 
  This work is supported in part by the U.S. DOE under DE-FG02-87ER40328 and 
  by the National Natural Science Foundation of China under Joint Research
  Grant No. 11128510.
\end{theacknowledgments}

\bibliographystyle{aipproc}

\end{document}